%% file: SOR_On_HPC_Languages.tex
\documentclass{doublecol-new}
\usepackage{epsfig}
\usepackage{multirow}
\usepackage{booktabs}
\usepackage{enumerate}
\usepackage{multicol}
\usepackage{amsmath}
\usepackage{color}
\usepackage{stfloats}

\theoremstyle{TH}{

}

\theoremstyle{THrm}{

}

\theoremstyle{THhit}{

}

\usepackage{flushend}
\usepackage[square, comma, sort&compress]{natbib}

\usepackage{epsfig}

\usepackage{subfigure}
\usepackage{graphicx}

\usepackage{amsmath}
\usepackage[lined,ruled,commentsnumbered,linesnumbered]{algorithm2e}

\usepackage{fancyvrb}
\usepackage{cprotect}

\usepackage{fancyheadings}
\makeatletter
\def\theequation{\arabic{equation}}
\makeatother
\usepackage{url}
\makeatletter
\def\url@leostyle{%
  \@ifundefined{selectfont}{\def\UrlFont{\sf}}{\def\UrlFont{\small\ttfamily}}}
\makeatother
\begin{document}%

\thispagestyle{plain}

\setcounter{page}{1}

\LRH{S. Mittal  }
\RRH{}

\VOL{x}

\ISSUE{x}

\PUBYEAR{x}

\title{A Study of Successive Over-relaxation Method Parallelization Over Modern HPC Languages}

 \authorA{Sparsh Mittal}
  \affA{Future Technologies Group\\
   Oak Ridge National Laboratory (ORNL)\\
   Oak Rdge, TN, USA\\
   Email: sparsh0mittal@gmail.com
 }

\input{abstract}

\KEYWORD{Successive over-relaxation (SOR), Chapel programming language, Go programming language (golang), D programming language (dlang), multithreading, high-performance computing (HPC), Parallelization.}

\REF{to this paper should be made as follows: Mittal, S. `A Study of Successive Over-relaxation Method Parallelization Over Modern HPC Languages', Int. J. of High Performance Computing and Networking, 2014.}

\begin{bio}
 Sparsh Mittal received the B.Tech. degree in electronics and communications engineering from IIT, Roorkee, India and the Ph.D. degree in computer engineering from Iowa State University, Ames, IA, USA. He is currently working as a Post-Doctoral Research Associate at Oak Rige National Laboratory (ORNL), TN, USA. In his B.Tech. degree, he was the graduating topper of the batch and his major project was awarded institute silver medal. He was awarded scholarship and fellowship from IIT Roorkee and Iowa State University. His research interests include high-performance computing, multicore system architecture and graphics processing units (GPUs). 
\break
\end{bio}

\maketitle
 \fancyhf{} 
\renewcommand{\headrulewidth}{0pt}
\thispagestyle{fancy}
\lfoot{Accepted in Int. J. of High Performance Computing and Networking, 2014.}
\vfill
\pagebreak

\noindent\parbox{42.4pc}{\leftskip 12.75pc\NINE .\vs{12}}



\input{intro}

\input{background}
\input{methods}

\input{features}

\input{algorithmCode}

\input{results}

\input{conclusion}

\bibliographystyle{IEEETR}

\bibliography{References}

%

\end{document}

%% file: abstract.tex
\begin{abstract}

Successive over-relaxation (SOR) is a computationally intensive, yet extremely important iterative solver for solving linear systems. Due to recent trends of exponential growth in amount of data generated and increasing problem sizes, serial platforms have proved to be insufficient in providing the required computational power. In this paper, we present  parallel implementations of red-black SOR method using three modern programming languages namely Chapel, D and Go. We employ SOR method for solving 2D steady-state heat conduction problem. We discuss the optimizations incorporated and the features of these languages which are crucial  for improving the program performance. Experiments have been performed using 2, 4, and 8 threads and performance results are compared with serial execution. The analysis of results provides important insights into working of SOR method.   

\end{abstract}



%% file: intro.tex
\section{Introduction}

Successive over-relaxation (SOR) is one of the most important method for solution of large linear systems (\cite{young2003iterative,hadjidimos2000successive,adams1986sor}). It has applications in CFD (computational fluid dynamics), mathematical programming (\cite{cryer1971solution}), medical analysis (\cite{ha2010multi}) and machine learning (\cite{mangasarian1999successive}) etc. The example of applications of SOR in CFD include study of steady heat conduction, turbulent flows, boundary layer flows or chemically reacting flows. For this reason, SOR method is important for both researchers and business policymakers.   

Due to recent trends of exponential growth in amount of data generated (\cite{mittal2011versatile,MitGup2013_QA}) and increasing problem sizes, serial platforms have proved to be insufficient in providing the required computational power. Hence, parallelization of computation intensive problems has become essential. Moreover, as the chip power-budget considerations restrict processor frequency-scaling, processor designers have focused on using tens of cores on a single chip to achieve high-performance and future processors are expected to have hundreds of cores. Thus, high-performance computing approach is expected to be even more useful for future systems.

In this paper, we present parallel implementations of SOR method 
using three concurrent programming languages, namely Chapel (from Cray Inc. (\cite{chamberlain2007parallel}),  D (from Digital Mars (\cite{alexandrescu2010d}))  and Go (from Google  (\cite{golangref})). The SOR method is used for solving 2D steady-state heat conduction problem. We discuss the relavant programming constructs of these languages and compare them to gain insights into their features which are crucial for improving the program performance. 
   
Experiments have been conducted with a square grid of dimension $4096 \times 4096$. Further, SOR method has been parallelized using 2, 4 and 8 threads using all the three languages and their performance has been compared with the serial execution. The analysis of results provides important insights into working of SOR method. The results also highlight the importance of using high-performance computing approach for obtaining solution of grid with large dimensions.

The rest of the paper is organized as follows. Section \ref{sec:relatedwork} presents a background on Chapel, D and Go languages and SOR method. Section \ref{sec:method} presents the algorithm, optimizations,  implementation details and salient features of our approach. Section \ref{sec:results} discusses the performance results. Finally, section \ref{sec:conclusion} concludes this paper.

%

%% file: background.tex
\section{Background and Related Work}\label{sec:relatedwork}
In this section, we briefly review the SOR method and programming features of Chapel, D and Go.

\subsection{Successive Over Relaxation (SOR) Method}
For solution of partial differential equations, both direct and iterative solvers have been used. The direct solvers are susceptible to round-off errors; while iterative solvers provide the opportunity to achieve desired accuracy by trading-off the speed. Moreover, the iterative solvers are generally more memory-efficient than the direct solvers and hence are  especially useful for solving large-sized problems. For this reason, we focus on iterative solvers in this paper. The examples of iterative solvers include Jacobi method, Gauss-Siedel (GS) method and SOR method. 
 
 The Jacobi method is a well-known iterative solver method which computes the value for iteration $k$ based on values from iteration $k-1$. The Jacobi method is easily parallelized, however, its slow convergence rate prevents its use for any real-life application. GS method partially makes use of new values to reach to convergence faster. SOR is an extrapolation of the GS method which works by using weighted version of previous and computed GS iterate to accelerate its convergence. 
 
\begin{equation}\label{eq:sormaineq}
X_k = \omega \overline{X_k} + (1-\omega) X_{k-1}
\end{equation}
Here $\overline{X_k}$ is the $k$-th Gauss-Siedel iterate and $0 < \omega < 2$ is the extrapolation factor.  By choosing a suitable value of $\omega$, the convergence rate of SOR can be improved. GS method is a special case of SOR for $\omega = 1$. Since methods to improve the speed of convergence is outside the scope of this work, we use a single value of $\omega$ as $0.376$.

The simulation is performed till the largest value of $|X_{k}(i,j)-X_{k-1}(i,j)|$ (i.e. difference between successive values of $X_k$ ) for any grid point ($i,j$) remains greater than $\epsilon$, where $\epsilon$ shows the numerical tolerance. By choosing a suitable value of $\epsilon$, a trade-off can be exercised between simulation speed and accuracy.

\subsection{ Chapel, D and Go Programming Languages}
HPC is a promising approach for accelerating computational-intensive applications, such as multimedia processing (\cite{GupMit08_MIMO,pande2009baywave}), atmospheric simulations (\cite{yang2013peta}), processor simulations (\cite{MitZha12_Simulation}),  medical imaging (\cite{liu2009real}), bioinformatics (\cite{zhang2011efficient}) etc. In recent years, HPC has been widely used by researchers (\cite{gupta2008guaranteed,ref13,mittal_sysgen,raju2012high}). Several parallel programming languages have been used such as D, Go, Chapel, Java, OpenMP, and X10 (\cite{charles2005x10}) etc.  These languages facilitate writing large-sized programs and provide programming constructs to express parallelism present in the program.  

In this paper, we use Chapel, D and Go to accelerate SOR method. These languages have some similarities along with some unique features. These languages provide concurrent programming facility as part of the language itself. Neither of them require VM (virtual machine), rather they are statically compiled. Chapel aims at improving the performance, programmability, portability, and robustness of high-end processors, while also facilitating parallel programming on commodity clusters or desktop multicore systems. Both D and Go are system programming languages and provide automatic garbage-collection, and faster compilation speed than C/C++ (\cite{hundt2011loop}). Chapel does not provide automatic garbage-collection.  
 

These languages also have some important differences. Relative to each other, Go aims more for simplicity and faster development, D aims for providing more features and Chapel aims for higher performance. Go does not have classes and only uses structs and interfaces; Chapel has classes and records, while D has classes and structs. Further, unlike D and Chapel, Go does not provide operator over-loading.  The differences in methods used for parallel programming are discussed in Section \ref{sec:method}.

\subsection{HPC Techniques for SOR Method}
Recently, several researchers have used GPU (graphics processing unit) to accelerate SOR method by offloading the memory intensive computations to GPU (\cite{di2010toward,itu2011gpu,konstantinidis2012graphics}). Compared to these, our implementation does not require a special-purpose hardware. In contrast to CPUs, GPUs have much smaller onboard memory and the bandwidth of the bus connecting CPU memory to GPU memory subsystem has limited bandwidth. Hence, for applications which require large working sets and number of iterations, the overhead of data-transfer between GPU and CPU presents a severe performance bottleneck (\cite{datta2008stencil}). Moreover, use of GPUs also entails the overhead of porting the legacy code to graphics hardware, which may be economically infeasible in many scenarios.




%
%
%
%



%% file: methods.tex
\section{Methodology}\label{sec:method}

\subsection{Parallelization of SOR Method}

In literature several parallel version of SOR have been proposed such as red-black SOR, multi-color SOR (\cite{adams1982multi}),  block-parallel SOR (\cite{xie2006new}). In this paper, we use red-black SOR method.

\begin{figure}[htp]
\centering
 \includegraphics [scale=0.65] {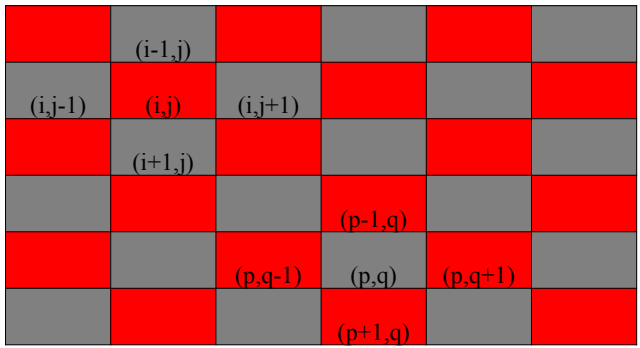}
\caption{The checkerboard configuration for SOR. Note that all red cells have black cells as their four neighbors and vice versa. }
\label{fig:sor_redblack}
\end{figure}

Red-black SOR divides the grid into a chessboard of red and black cells, as shown in Figure \ref{fig:sor_redblack}. For a given row value $i$ and column value $j$, a red cell is one for which $(i+j)$ is even and a black cell is one for which $(i+j)$ is odd. Clearly, all red cells have black cells as their neighbours and vice-versa. 
 
 \begin{figure*}[htp]
\centering
\cprotect\fbox{
\begingroup
    \fontsize{7.5pt}{9.5pt}\selectfont
 \begin{minipage}{.5\textwidth}
 \underline{\textbf{Code 1: Requires checking {\tt if} condition several times}}    
\begin{Verbatim}
for (i= 0; i < DIM; i++)
    for(j= 0; j< DIM; j++)
{
    if ( (i+j)%2 ==0)
      doProcessing()
}

\end{Verbatim}
\end{minipage}%
\begin{minipage}{.5\textwidth}
\underline{\textbf{Code 2: Avoids checking {\tt if} condition by restructring {\tt for} loop}}
\begin{Verbatim}
 for (i= 0; i < DIM; i+= 2)
    for(j= 0; j< DIM; j+= 2)
{    
      doProcessing()
} 
for (i= 1; i < DIM; i+= 2)
    for(j= 1; j< DIM; j+= 2)
{    
      doProcessing()
} 
\end{Verbatim}
\end{minipage} 
\endgroup
}
\caption{Optimization: Avoiding {\tt if} (branch statement) by restructring {\tt for} loop for red phase. Similar idea applies to black phase also. Here {\tt DIM} shows dimension of the grid. }
\label{fig:illustration}
\end{figure*} 
The red-black group identification strategy along with use of the five-point finite-difference stencil leads to uncoupling of the solution of Eq. \ref{eq:sormaineq} at interior cells such that the value at the red cells depends only on the value at the black cells and vice versa. Thus, red-black SOR divides the iteration in two steps, namely red phase and black phase.  In any iteration, first red cells can be updated and then for updating the black cells, the value just computed for red cells can be used. Clearly, such a strategy allows straightforward parallelization.


\subsection{Using Chapel, D and Go For Parallelization}
SOR method involves iterative computations, where threads executing the same code in parallel must all complete one phase (viz. red or black) of the iteration before moving on to the next phase (or iteration). To ensure this, a synchronization barrier is used which enables worker threads to wait until all the threads have all completed a phase before any thread continues. We now describe the parallelization approach and relavant programming constructs of Chapel, D and Go which enable us to achieve these functionalities. 
 
\subsubsection{Chapel Programming Language}
In Chapel language, we have utilized the task-parallel construct {\tt begin} along with synchronization construct {\tt sync}. Using {\tt begin}, the solver function is issued in asynchornous manner and using {\tt sync} statement, barrier synchronization is achieved.  

\subsubsection{D Programming Language}

In D language, we have utilized the functionality of {\tt std.parallelism} module. Each worker is started as a new {\tt task}. Using {\tt put} command, a {\tt task} is queued to the {\tt taskPool} for execution. The {\tt taskPool} encapsulates a task queue and executes the task by efficiently mapping them onto the threads. For barrier synchronization, {\tt yieldForce} function is used for each running task, and thus, the control waits till all the threads have finished execution. 

To allow different threads to access the global data, {\tt shared} qualifier is used to designate that a piece of data  is shared in different threads. Otherwise, by default, the data is local to each thread and hence, multiple threads cannot safely access it. For designating that a piece of data is constant and hence safe for concurrent reading, {\tt immutable} keyword is used.

\subsubsection{Go Programming Language}

In Go, we use {\textit Goroutines} to achieve concurrent programming. A function which is called with {\tt go} in its front is executed in its own goroutine. As an example, for the following code, 

\begin{verbatim}
go function1()  
function2()
\end{verbatim}
both {\tt function1} and {\tt function2} run concurrently. Goroutines are lightweight and are multiplexed onto a set of threads. When a goroutine is blocked, the run-time automatically moves other goroutines on the same operating system thread to a different, runnable thread to allow maximum utilization of the resources.

 The maximum number of processors to be used are specified using {\tt GOMAXPROCS} function from {\tt runtime} package. For implementing barrier synchronization, we used {\tt WaitGroup} variable. Using {\tt Add} function, the number of goroutines to wait for is specified and each such goroutine issues {\tt Done} function to signal completion. When all goroutines complete,  the barrier is released.

%% file: features.tex
\subsection{Optimizations Incorporated} \label{sec:features}
To achieve high performance, we have applied several optimizations to both serial and parallel versions. We now discuss these optimizations briefly and then present the algorithm in the next section. 

\textbf{Code restructring to avoid branch misprediction penalty:} Modern processors use long pipelines to allow instruction-level parallelism (ILP), however, this also increases the penalty of branch mispredictions. Branch prediction is required for predicting the outcome of conditional statements like {\tt if}.  To reduce branch misprediction penalty, we have structured the {\tt for}-loops in a manner that {\tt if}-condition checking is minimized, as shown in Figure \ref{fig:illustration}. This also allows the compiler to apply optimization techniques such as loop-unrolling and vectorization, if applicable. Moreover, since branch conditions are removed, the runtime performance is significantly improved.

\textbf {Minimizing serial execution bottleneck: } In SOR method, checking for convergence requires finding the maximum absolute error for all the cells. This would require comparing the error value at all the cells to a single {\tt maxChange} value (see Algorithm \ref{alg:parallelsor}). This represents a critical section and to avoid data race condition, mutex functionality might be required. To avoid it, convergence check is done in a serial manner (see Algorithm \ref{alg:parallelsor}). This increases the storage overhead slightly, but it also provides performance improvement.

\textbf{Choice of granularity of convergence check:} In SOR method, since generally convergence is reached after thousands of iterations, testing for convergence  at the end of each iteration may lead to extra computations (e.g. comparisons).  To reduce the overhead of convergence test, we perform it only once after every {\tt K} iterations. The choice of {\tt K} presents a trade-off since a large value of {\tt K} may lead to performing extra iterations even after convergence is reached and a small value of {\tt K} may lead to increase overhead of convergence checking. We have heuristically chosen {\tt K} as 4000 to keep a balance between these two factors.

%% file: algorithmCode.tex
\begin{algorithm}[htbpH]
{
\caption{Parallel SOR Algorithm (see Section \ref{sec:mainalgo})}\label{alg:parallelsor}
 \fontsize{7.5pt}{9.5pt}\selectfont
\DontPrintSemicolon
\SetAlgoLined

\KwIn{Initial temperature profile, $P$ (number of workers) and $\omega$. }
\KwOut{Final temperature profile and whether SOR converged}
\textbf{Constants Used}: {\tt MaxIterations} (max number of iterations), {\tt K} (number of iterations after which convergence is checked) and  $\epsilon$ (tolerance) \;

\textbf{Variables Used}: {\tt gridData} and {\tt gridDataOld}: 2D arrays, {\tt hasConverged} = false,  {\tt maxChange}= 0.0  and {\tt shouldCheckConvergence} (whether to check for convergence in this iteration) = false\;
\BlankLine
Initialize the {\tt gridData} with initial temperature profile \;

\underline{\textbf{Algorithm for main routine}}\;

\ForEach{iteration {\tt iter} = 1 to {\tt MaxIterations} }
{
      
      \eIf{  {\tt iter} is a multiple of {\tt K} }
    {
          {\tt shouldCheckConvergence} = true \;
          Copy entire {\tt gridData} to {\tt gridDataOld} \;
    }
    {
           {\tt shouldCheckConvergence} = false \;
    }
    
      Call updateGridRed with $P$ workers in parallel \;
      Synchronize \;
    
      Call updateGridBlack with $P$ workers in parallel \;
      Synchronize \;
      
      \If{ {\tt shouldCheckConvergence} } 
    {
       {\tt maxChange =0} \;
         
        \ForEach {Cell $(i,j)$ in the grid} {
        {\tt maxChange } = Maximum($|gridData(i,j)-gridDataOld(i,j)|$, {\tt maxChange }) \;
       }
    
       \If {{\tt maxChange} $<$ $\epsilon$} {
       hasConverged = true \;
       break \; 
      }
    }

}
Print value of hasConverged. Return. \;

\BlankLine

\underline{\textbf{updateGridRed() for worker $p_j$}} \;
\ForEach{ Cell of red color given to worker $p_j$}
{
  Update {\tt gridData} using Eq. \ref{eq:sormaineq} \;  
}

\underline{\textbf{updateGridBlack() for worker $p_j$}} \;
\ForEach{ Cell of black color given to worker $p_j$}
{
  Update {\tt gridData} using Eq. \ref{eq:sormaineq} \;  
}

}
\end{algorithm}

%% file: results.tex
\begin{table*}[ht]\normalsize
\caption{Execution time and speedup values for different languages and different number of cores}
\label{tab:sor_results}
\centering
\begin{tabular}{|c||c|c|c||c|c|c|}
\hline
 & \multicolumn{3}{|c||}{Execution time in seconds}  & \multicolumn{3}{|c|}{Speedup relative to serial execution}\\ \hline
Threads      & Chapel & D & Go    & Chapel & D & Go\\ \hline
1 (Serial) & 7538 & 8609 & 10551 & - & - & - \\ \hline 
2 & 3977 & 4099 & 5204           &  1.90  & 2.10  & 2.03   \\ \hline
4 & 3139 & 3322 & 3834           &  2.40 & 2.59  & 2.75  \\ \hline
8 & 2824 & 3141 & 3052           &  2.67 & 2.74  & 3.46   \\\hline
\end{tabular}	
\end{table*}
 \subsection{Parallel SOR Algorithm}\label{sec:mainalgo}

Algorithm \ref{alg:parallelsor} shows the parallel SOR algorithm for the case of 2D steady-state heat conduction problem. Initially, the temperature at north boundary is assumed to be 1.0 unit, while at all other boundaries, it is assumed to be 0.

The algorithm proceeds as follows. Initially, the grid is initialized. The main routine runs for a maximum of {\tt MaxIterations} number of iterations. In each iteration, red and black phases spawn $P$ workers (e.g. threads or goroutines) which need to be synchronized at the end of the phase. Each worker updates the cells of the given color allocated to it. Convergence check is performed only when  {\tt shouldCheckConvergence} is true, which happens after every {\tt K} iterations. The serial version of SOR algorithm follows along the same lines and has been omitted for the sake of brevity.  

\section{Experimental Platform and Results} \label{sec:results}

  To gain highest performance in the final run, we compiled Chapel code with {\tt --fast} flag (which also turns ON the {\tt -O3} flag for the compilation of the back-end C code produced by the Chapel compiler ) and D code with {\tt -inline -O -release} flags. For Go code, no suitable optimization flag could be found.  The values of constants are taken as $\epsilon$ = 0.00001, {\tt MaxIterations} = 50000 and {\tt K} = 4000. The dimension of the grid is $4096\times 4096$. For each of the three languages, we wrote both serial and parallel programs. We compare the performance scaling of these languages by comparing the execution time of parallel programs with the serial program written in the same language. Table \ref{tab:sor_results} presents these execution time values. It also presents the speedup values, where speedup $S_P$ for $P$ threads is defined as
  
  \begin{equation}
  S_P = \dfrac{T_1}{T_P}
  \end{equation}
  
  Here $T_1$ and $T_P$ refer to the execution time using 1 (serial) and $P$ threads, respectively for the same language.

We now analyze these results. For the SOR program, out of the three languages tested, Chapel language gives the highest performance (Note that the above mentioned results should not be taken to conclude about the performance of these languages for all possible programs, since a different program and coding style might produce different results). For small number of threads (e.g. 2), the performance scales nearly linearly (A slight variation can be attributed to load on the host machine), however, for large number of threads (e.g. 8), the performance does not scale linearly. This is due to the fact that SOR program involves multiple iterations and  phases; and the synchronization required after each phase creates serialization bottleneck due to which multiple threads do not progress independently. If the number of iterations required for convergence is $L$, then synchronization is required for $2\times L$ times. Moreover, with the increasing number of cores, although the processing power increases, the other resources such as cache, memory bandwidth etc. do not increase linearly and hence, the program performance does not scale linearly.  



%% file: conclusion.tex
\section{Conclusion and Future Work}\label{sec:conclusion}
In this paper, we have highlighted the importance of using HPC approach \cite{mittal_simplex} for accelerating solution of SOR method. We presented parallel implementations of SOR method for solving 2D heat transfer problem. We discussed the features of Chapel, D and Go, along with their functions which are important for achieving parallel implementation of SOR method. Experimental results have been conducted using 2, 4 and 8 threads and analysis of results provides important insights into SOR method. Our future work will focus on solving the SOR problem for 3D grids. We also plan to study parallelization of other computation intensive problems. 
